# Multicore Applications in Real Time Systems

[1]Vaidehi M. [2] T.R.Gopalakrishnan Nair

[1]Research Associate, vaidehidm@yahoo.co.in
[2]Director- Research & Industry, trgnair@ieee.org
[1,2] Research & Industry Incubation Center, Dayananda Sagar Institutions, Bangalore, India

**Abstract** - Microprocessor roadmaps clearly show a trend towards multiple core CPUs. Modern operating systems already make use of these CPU architectures by distributing tasks between processing cores thereby increasing system performance. This review article highlights a brief introduction of what a multicore system is, the various methods adopted to program these systems and also the industrial application of these high speed systems.

Index terms - *Automatic load balancing, Asymetric Multi Processing (AMP), Bound Multi Processing (BMP) Real time systems Symetric Multi Processing (SMP), Scaling.*

## I. Introduction

Requirement of high speed processors is vital in real time systems. As the cost of high-speed processors has reduced, it has paved the way for solving real time system demands more effectively. Today, multicore systems are reviewed in real time embedded systems.

Few of the most prospective areas where multicore systems can be applied are wireless network applications, cognitive systems, image recognition units, biomedical systems and automobiles.

### A. Multicore Processors

A multi-core processor is an integrated circuit to which two or more processors have been attached for enhanced performance, reduced power consumption and efficient simultaneous processing of multiple tasks. A dual core set-up is somewhat comparable to having multiple, separate processors installed in the same computer, but because the two processors are actually plugged into the same socket, the connection between them is faster. Ideally, a dual core or a quad core processor is nearly twice as powerful as a single core processor [1]. In practice, performance gains are said to be about fifty percent. A dual core processor is likely to be about one-and-a-half times as powerful as a single core processor. Multi-core processing is a growing industry trend as single core processors rapidly reach the physical limits of possible complexity and speed. Companies that have produced or are working on multi-core products include AMD, ARM, Broadcom, Intel, and VIA.

## II. Different Ways of Processing a Multicore System

Asymmetric multiprocessing and symmetric multiprocessing are terms used to describe a Real-Time operating system and hardware architecture of the device. Here we discuss the high level advantages and disadvantages of each in a multicore environment and why the need for a simple communication scheme exists no matter which architecture is used. **Symetric Multi Processing** (SMP) has a single image of the (Real Time Operating System) RTOS shared by many cores, they handle load balance between the cores easily. Another benefit is that all the data is available for all the cores and tasks, hence any external communication mechanism between cores is not needed. All communications are handled within the SMP RTOS. The major disadvantage of SMP systems is the inability for the SMP RTOS to operate in a heterogeneous environment.

AMP RTOS differs from SMPs in that instead of sharing cores with one image, AMPs have a number of images per core. The main advantage of an AMP is that it can be used in a heterogeneous environment. AMP systems with large a number of cores can face problems with scalability. AMP systems need some type of mechanism to communicate among the cores to realize the full potential of multicore systems.

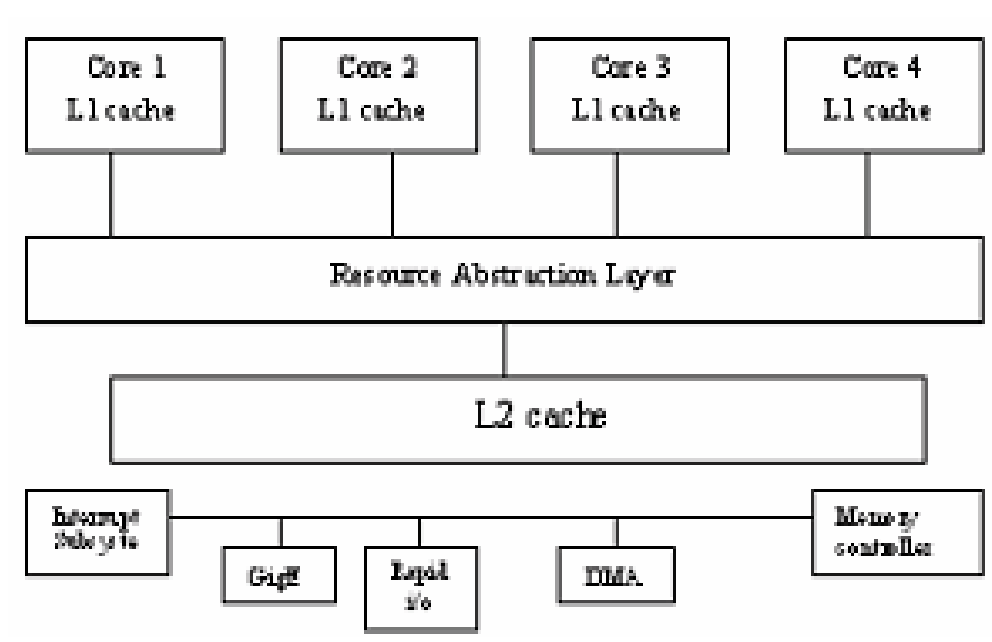

**Figure 1** Anatomy of a typical multi-core processor [6]

Bounded multiprocessors offer the benefits of SMP's transparent resource management, but give designers the ability to lock software tasks to specific cores.

This method eliminates the cache thrashing that can reduce performance in an SMP system allowing applications that share the same data set to run exclusively.

The advent of multicore processor is exerting greater demands on applications and programmers who build them. In a symmetrical multiprocessing model, certain mechanisms built into the RTOS can greatly ease that burden. Multiprocessors can be configured in a variety of forms, from loosely coupled computing grids that use the Internet for communication to tightly coupled Symmetric Multiprocessors (SMP). As shown in Figure 1, all processors in an SMP system have access to the same physical memory. Since they all have identical instruction sets as well, a process or thread can run on any processor from the same memory location. This is the key to adapting an application to SMP architecture.

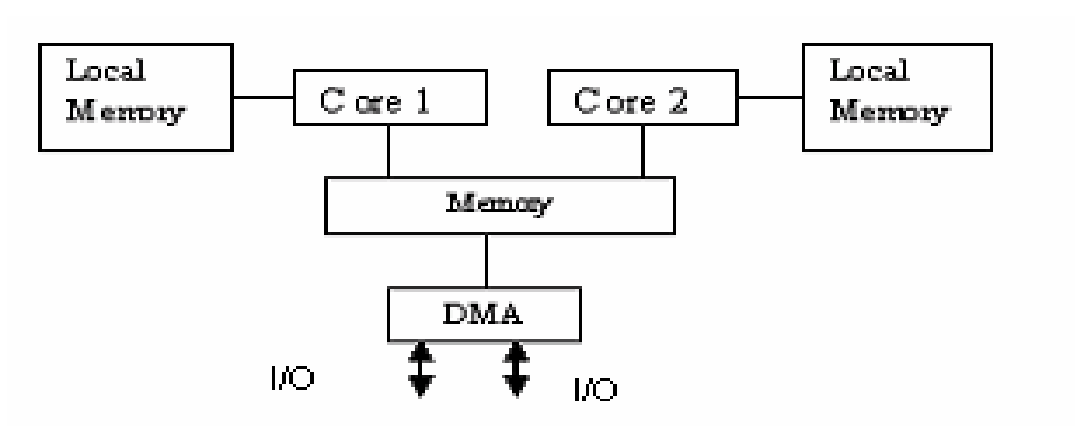

**Figure 2** In a simplified SMP architecture, multiple, identical processors have access to common memory and optionally also to private local memory that generally can be accessed with lower latency.

### A. Programming of the Multicore Systems

While multiprocessors offer exciting opportunities for power-efficient performance, achieving the goals of code reuse and fast development times are serious challenges. Traditional programming paradigms are single-processor-oriented, with logic that is not easily split among processors. Even when the workload is divided among multiple processors, the need for the programmer to coordinate work among processors can be demanding. Programming tools that assist in the mapping of applications to multiple processors are not generally available. However, there is a golden opportunity to use a software that lies between the application and the multiprocessor to help abstract the hardware for the application—the operating system. In embedded applications, this means the real-time operating system (RTOS) that is found in the most intelligent software has a special feature to manage multiple tasks or threads of operation.

The RTOS can help an embedded application in a number of ways. Typically, the RTOS is used to handle interrupts and manage the scheduling of application threads. A thread is a sequence of logically complete codes that perform a particular role. Embedded real-time applications generally are made up of multiple threads; each performing its intended function, scheduled according to priorities or in response to external events (interrupts). In a multiprocessor system, in addition to these services, the RTOS can also distribute processing across all processors. This frees the application from deciding what should be programmed to run where.

The RTOS can also respond to variations in processing load. This enables all processors to be utilized during periods of maximum demand without requiring explicit assignment by the application. To achieve minimum power consumption, the RTOS can adjust clock frequency or shut down individual processors to conserve power during periods of light demand. By using an RTOS to interface with the underlying SMP architecture, applications are kept hardware-agnostic and this enables a "write once, run anywhere" approach that maximizes code reuse across multiple architectures, whether they are single processor or multiple processor. These RTOS capabilities combine to make the developer's job easier. And that is the key to enabling shorter and less risky development projects with faster time-to-market, while still reaping the benefits of a complex multiprocessor architecture

### B. Scaling Partitions on Multi-core Processors

Till date, partitioning has been used almost exclusively in single-processor environments. However, with the growing proliferation of multi-core processors, developers now need a way to implement partitions across two, four, eight, or more processing cores.

Therein lies the challenge. In a single-processor environment, the RTOS scheduler allocates CPU capacity to each partition. Ideally, the RTOS scheduler can simply extend this concept across all the processing cores in a multi-core system. Unfortunately, many RTOSs do not provide this capability, largely because of their limited multiprocessing capabilities.

An RTOS may support one or more of the following multiprocessing models:

- Asymmetric multiprocessing (AMP) — treats each core as a discrete CPU. A separate instance of the RTOS runs on each core, forcing the developer to statically configure memory, interrupts, and other shared system resources. Applications running on a given core can use only the resources that have been statically configured for that core.
- Symmetric Multiprocessing (SMP) — a single instance of the RTOS manages all the cores. The RTOS transparently manages shared system resources, allowing them to be used by any application running on any core. In addition, the RTOS can dynamically schedule any process on any core, enabling full CPU utilization. This approach not only simplifies development, but also offers greater flexibility when using secure partitions in a multi-core environment.
- Bound Multiprocessing (BMP) — Extends SMP by allowing the developer to bind any process (and all of its associated threads) to a specific core. This approach combines the developer control of AMP with the simplified programming and greater scalability of SMP.

By its very nature, AMP cannot address processing requirements that extend across multiple cores. Thus, any partition in AMP is limited to a portion of a single core, upto 100%. In SMP and BMP, on the other hand, the RTOS has an overall system view, allowing it to use the entire CPU capacity of the multi-core processor (*i.e.* all the cores) for partitioning. System designers can, as a result, flexibly map partitions onto a number of cores in whatever manner is dictated by system requirements — and independently of processor boundaries. For instance, in Figure 2, Secure Partition 1 spans across two CPU cores while the other partitions run on single cores.

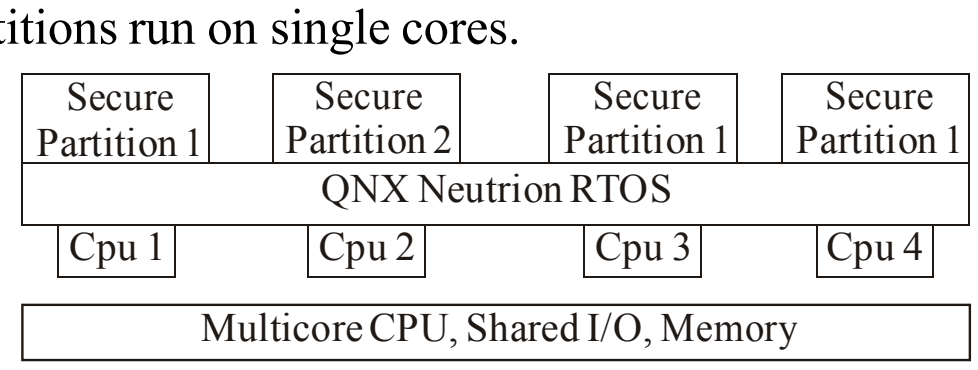

**Figure 3** In the symmetric and bound multiprocessing and bound multiprocessing Models, partitions can flexibly span across multiple cores.

### C. SMP RTOS Goals

While SMP architectures have been used in enterprise computing for a long time, and SMP operating systems abound, SMPs are relatively new to embedded computing. Currently, Linux supports SMP architectures, and can be adapted to support many embedded applications. But for applications that demand more real-time responsiveness than Linux is able to provide, or that require a smaller memory footprint than that required by Linux, an RTOS would have to be adapted to manage and exploit the facilities of an SMP

A typical application domain that could be addressed by an SMP RTOS is that of handling a stream of data where there is a requirement for substantial real time processing of the data in each packet in the stream. This is the case with streaming audio, video or multimedia data that requires compression/decompression, encryption/de-encryption, rendering, filtering, scaling and other CPU-intensive processing. In a typical system—a cell phone for example—data might be placed via DMA into buffers sized to suit the target display or other system characteristics. Once a buffer is full, an interrupt is generated to alert the application, and that buffer is targeted for processing while data streams into a new buffer. This particular model describes a data flow that is continuous, but varying in intensity. In such a model, real-time response to data arrival is important and overall throughput is the goal, but programming convenience is the key.

An RTOS that is designed to manage and utilize the resources of an SMP system can be beneficial in such applications. The RTOS must be adapted to run on the SMP architecture. This means that it must be ported to work with the SMP and the underlying processor architecture. Development tools must exist to support the RTOS for that architecture.

Next, the RTOS must enable utilization of both the processors, enabling both the processors to work on data simultaneously, and manage operation of multiple instruction streams on independent processors. Ideally, some automation of processor utilization can be provided to avoid forcing the programmer to manage load balancing or processor switching from within the application. The RTOS can automatically, transparently and dynamically determine when to assign certainprocessing to certain processor resources.

The RTOS must also handle interprocessor communication, allowing the programmer to focus more on the algorithm rather than the architecture. The RTOS must enable synchronization among processors, and provide a mechanism for interprocessor communication (IPC).

If derived from an existing RTOS, the SMP RTOS should use and retain as much of that existing RTOS as possible, which will help provide easy migration of legacy (single processor) applications to the SMP. A familiar RTOS makes it easier to learn to manage an SMP.

## III. Automatic Load Balancing

One of the most intriguing aspects of delivering a performance increase from an SMP is the ability to do so without requiring programmer intervention. This not only makes it easier for the developer but, more importantly, it makes it possible to use legacy application code in an SMP system without modification.

One method of achieving programming transparency in a multiprocessor system is to assign individual threads to run on specific processors based on the availability of the processor. This way, the processing load can be shared among processors with work automatically assigned to a free processor. The RTOS must determine whether a processor is free and if it is, then a thread can be run on that processor even though the other processors may already be running other threads. This enables a more complete utilization of resources, yet remain transparent to the number of processors and to the programmer, and enable legacy code to be used intact.

In order to utilize such an approach, the developer must set up multiple, identical threads, allocate threads to process portions of the data stream, and set equal priorities. Priorities are important to consider because the RTOS scheduler is designed to maintain priority execution of all threads, such that higher priority threads get executed before lower priority threads. This way, threads can safely assume that while they are executing, no lower priority thread can be executing. The RTOS must preserve this rule even in the case of an SMP, or the underlying logic upon which a legacy application might be based could falter, and the application may not perform as intended.

Priority-based, pre-emptive scheduling uses multiple cores to run threads that are ready. The scheduler automatically runs threads on available cores. A thread that is "READY" can run on processor-n if and only if it is of the same priority as the Thread(s) running on processor(s)-(n-1). After initialization, the RTOS scheduler determines the highest priority thread that is READY to run. It sets the context for that thread, and runs the thread on processor-1. The scheduler determines if an additional thread of equal priority is also READY. If so, that thread is run on processor-2, and so on. If no additional threads are ready to run, the scheduler goes idle awaiting an external event or service request, such as an interrupt causing pre-emption, or a thread resume, sleep, relinquish or exit.

Pre-emption occurs when a thread is made READY to run while a lower priority thread is already running [2]. In this event, the lower priority thread is suspended (context saved), the higher priority thread is started (context restored or initialized), and any lower priority threads on other processors are suspended. This is critical to maintain the priority order of executing threads.

Within this automatic load-balancing approach to managing the resources of an SMP like ARM's MP core, additional features are beneficial to overall performance. One processor can be made responsible for all external interrupt handling (this does not include interprocessor interrupts needed for synchronization or communication). This leaves the other processor(s) with virtually zero overhead from interrupt handling, enabling it (them) to focus all of its (their) cycles on application processing, even during periods of intense interrupt activity that otherwise might degrade performance [3].

As an example, consider Figure 4, which shows a two-processor system that is intended to handle a continuous stream of incoming data, such as streaming video. The data must be decompressed in real time. This is a typical data flow and processing model using an RTOS with automatic load-balancing support for an SMP.

Input is set up to fill Buffer 1 in memory, with an interrupt generated upon a buffer-full condition (or based on input of a specified number of bytes). As Buffer 1 reaches a full condition, Interrupt 1 is generated. In response, the ISR handling Interrupt 1 marks Thread 1 READY-TO-RUN, and the scheduler runs Thread 1 on Processor 1. Simultaneously, data is directed to Buffer 2, and processor 2 remains idle for the moment.

Then, as more data arrives while Thread 1 is active, Buffer 2 fills up, generating Interrupt 2. The ISR handling Interrupt 2 marks Thread 2 READY-TO-RUN, and the scheduler runs Thread 2 on Processor 2, while Thread 1 continues to run on Processor 1.

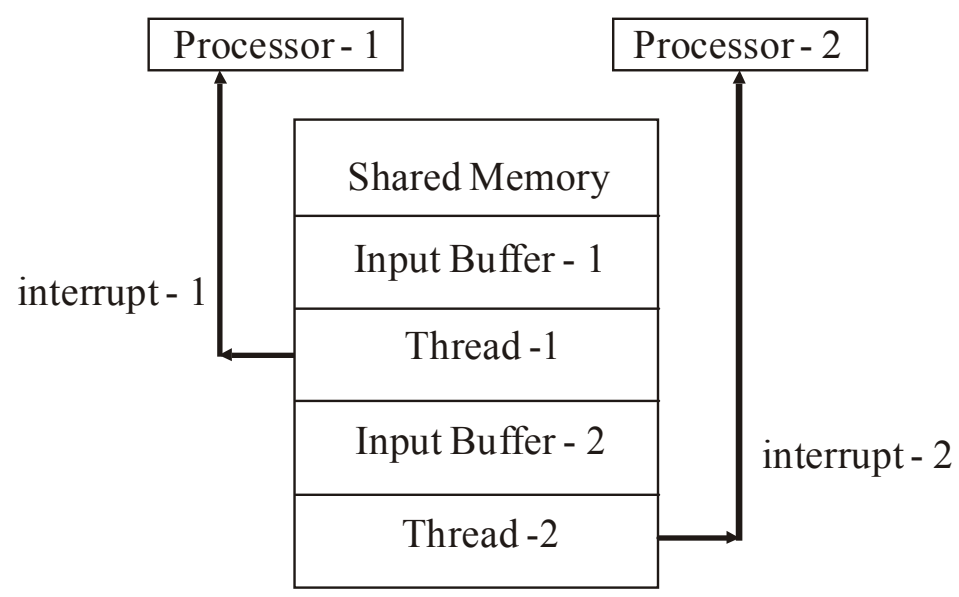

**Figure 4** Two processors are allocated by the RTOS to work on separate data sets as they arrive in the system.

An RTOS that can manage an SMP system can provide increased performance compared to a single processor system, while minimizing demand on the development of applications. This delivers the benefit of greater performance with little effort, making SMP architectures very attractive for embedded applications requiring high performance, but with stringent demands on time-to-market [4] [5].

### A. Choosing between AMP, SMP and BMP

The choice between AMP, SMP, and BMP depends on the problem you are trying to solve:

AMP works well with legacy applications, but has limited scalability beyond two CPUs.

SMP offers transparent resource management, but software that has not been properly designed for concurrency might have problems.

BMP offers many of the same benefits as SMP, but guarantees that uniprocessor applications will behave correctly, greatly simplifying the migration of legacy software. As the following Table 1 illustrates, the flexibility to choose from any of these models lets you strike the optimal balance between performance, scalability and ease of migration.

## IV. Industrial Applications of Multicore Processors

Various industrial areas such as military, transportation, automation and control etc work with multicore systems because of their intensive advantages. As mentioned earlier, application of multicore systems improve upon the performance of the system with high processing speed, less power consumption etc. Below, it is mentioned in brief, how multicore systems are applicable in the area of military.

In order to advance its tactical capability, the military requires more computing functionality in compact battlefield systems. New multicore processors used in embedded devices provide more functionality, increase system performance and run at lower temperatures. The multicore operating environment introduces a new software paradigm where general purpose and real time operating systems and applications need to run concurrently. For example, a multicore military application could have an unmanned mobile device or a robot providing live video of unexploded ordinance to an operator with an RTOS. The mobile device can collect additional data for processing against databases stored on a general purpose database on general purpose operating systems like Windows or Linux. An operator would control an unmanned robot to disarm the suspected device. In this case, both real time and general purpose OSs are required to complete the task.

**Table 1** Scheduling of the Tasks in Bound Multicore Processor

| Feature | SMP | BMP | AMP |
|---|---|---|---|
| Seamless resource sharing | Yes | Yes | -- |
| Scalable beyond dual CPU | Yes | Yes | Limited |
| Legacy application operation | In most cases | Yes | Yes |
| Mixed OS environment (e.g. Neutrino and Linux) | -- | -- | Yes |
| Dedicated processor by function | -- | Yes | Yes |
| Intercore messaging | Fast (OS primitives) | Fast (OS primitives) | Slower (application) |
| Thread synchronization between CPUs | Yes | Yes | -- |
| Load balancing | Yes | Yes | -- |
| System-wide debugging and optimization | Yes | Yes | -- |

## V. Conclusion

Multicore architectures create new challenges for the application programmer. Fortunately, good planning and utilization of a well-designed RTOS can make these challenges manageable. In fact, a good RTOS allows the programmer to write much of the code in the same way as it would be written on a single-core processor. The key is to understand the special requirement of the multi-core environment and to know how to use the RTOS to meet these requirements.

**Table 2** Multicore real-time and general purpose embedded system applications

| Real Time Application | Windows and Linux General Purpose Application |
|---|---|
| Medical | Database |
| Scientific | Graphic display |
| Transportation | Internet |
| Automation and Control | Imaging |
| Military | Media broad cast |
| Multimedia | Web design |